\documentclass[5p]{elsarticle}

\usepackage{graphics}
\usepackage{eufrak}

\newcommand{\be}{\begin{equation}}
\newcommand{\ee}[1]{\label{#1} \end{equation}}
\newcommand{\ba}{\begin{eqnarray}}
\newcommand{\ea}[1]{\label{#1} \end{eqnarray}}

\newcommand{\exv}[1]{ \langle {#1} \rangle}

\begin{document}

\title{A q-parameter bound for particle spectra based on black hole 
thermodynamics with R\'enyi-entropy}

\author[rmki]{Tam\'as S.~Bir\'o}
\ead{biro.tamas@wigner.mta.hu}
\author[rmki,cmat]{Viktor G.~Czinner}
\ead{czinner.viktor@wigner.mta.hu}
\address[rmki]{HAS Wigner Research Centre for Physics, 
Institute for Particle and Nuclear Physics,
H-1525 Budapest, P.O.Box 49, Hungary}
\address[cmat]{Centro de Matem\'atica, Universidade do Minho, 
Campus de Gualtar, 4710-057 Braga, Portugal}

\date{\today}

\begin{abstract}
{By regarding the Hawking-Bekenstein entropy of Schwarzschild black hole horizons 
as a non-extensive Tsallis entropy, its formal logarithm, the R\'enyi entropy, is 
considered. The resulting temperature -- horizon-radius relation has the same 
form as the one obtained from a 3+1-dimensional black hole in anti-de Sitter space 
using the original entropy formula. In both cases the temperature has a minimum.
A semi-classical estimate of the horizon radius at this minimum leads to a Bekenstein 
bound for the $q$-parameter in the R\'enyi entropy of micro black holes, ($q \ge 1+2/\pi^2$), 
which is surprisingly close to fitted $q$-parameters of cosmic ray spectra and 
power-law distribution of quarks coalescing to hadrons in high energy accelerator 
experiments.}
\end{abstract}


\maketitle

\section{Introduction}

Non-extensive thermodynamics \cite{NEXT1} aims at describing dynamically and statistically 
entangled systems. It is noteworthy that already the \emph{ideal gas}, a constant heat capacity 
system, shows a power-law tailed distribution of the energy in the microcanonical treatment.
In fact the Tsallis entropy can be viewed as
\be
 S_{{\rm Tsallis}} = K(S_{{\rm Renyi}}) = \sum_i p_i K(-\ln p_i),
\ee{Tsallis_RENYI}
with
\be
K(S) = \frac{1}{a} \left( e^{aS} - 1\right).
\ee{KFUNC}

In general $a$ is a real parameter, for an ideal gas in particular,
and $a=1/C_0$ with $C_0$ being the constant, positive heat-capacity \cite{IDEAL1,IDEAL2}.
$K(S)$ is additive for the finite ideal gas, while $S$ is not. 
In the $C_0\rightarrow\infty$ limit $a=0$ and the Boltzmann-Gibbs formula follows.
It is, in general, a better approximation to consider a constant, but finite value
for the parameter $a$.

By these definitions the familiar Boltzmann entropy is given as the expectation
value of the 'surprise' measure, $\mathfrak{s}=-\ln p_i$:
\be
S_B = - \sum p_i \ln p_i = \exv{\mathfrak{s}},
\ee{SURPRISE}
while the R\'enyi entropy is the logarithm of the characteristic function:
\be
S_R = \frac{1}{a} \ln \sum p_i^{1-a} = \frac{1}{a} \ln \exv{e^{a\mathfrak{s}}}.
\ee{LOGCHAR}
The latter generates an infinite series based on the central moments (statistical correlations):
\be
S_R = \exv{\mathfrak{s}} + \frac{a}{2!} \delta_2\mathfrak{s} + \frac{a^2}{3!} \delta_3\mathfrak{s} + \ldots
\ee{SR_AS_CMOM} 
Usually all central moments scale with the volume $V$ of the system, and interpreting
$a=1/C_0$ as discussed and cited above, it scales like $1/V$. This way the R\'enyi entropy
converges to the Boltzmannian formula as fast as $V^{2-n}/n!$.

The Tsallis entropy, on the other hand, is given as
\be
S_T = K(S_R) = \frac{1}{a} \left(\exv{e^{a\mathfrak{s}}}-1 \right),
\ee{CHAR_TSALLIS}
and it generates a series based on the direct statistical moments, with the generic term
being $a^{n-1}\exv{\mathfrak{s}^n}/n!$. All terms in this series are of the same order of $V$.
So for fast convergence to the classical results the R\'enyi entropy is preferable,
although, it is only the Tsallis formula which is linear in a direct expectation 
value based on the $p_i$ probabilities, offering in this way advantages  for transport 
theory and other linear response treatments.
Both formulas are stable close enough to the entropy maximum -- there is no difference
from this point of view in equilibrium calculations, which we shall pursue in this paper.

By using a non-Boltzmannian formula for the entropy, the experimentally observed power-law 
tailed particle spectra can be interpreted as reflecting a canonical ensemble in the non-extensive 
thermodynamics \cite{NEXT2,NEXT3,NEXT4,NEXT5}. The theoretical prediction of the power in 
single-particle spectra, related to the Tsallis parameter, $q=1+a$, is however a very difficult problem;
it requires a physical model of the heat-container present at the source
of particle emission. Simplified models, such as dual black holes, became
popular lately in this respect \cite{SON,Mat1,Mat2,GYULASSY}.

In earlier works we have fitted several hadron spectra observed at
RHIC \cite{PHENIX1,PHENIX2,PHENIX3,PHENIX4,STAR1,STAR2,STAR3} 
by assuming a blast wave picture and quark recombination \cite{SQM2008,FORMLOG}. 
These fits agreed with a common Tsallis parameter of $q \approx 1.2$ for the quark matter quite 
remarkably. According to the quark coalescence picture the corresponding value of $(q-1)$ for mesons
should be the half, and for baryons one third of this fit. 

A power-law tailed energy-distribution can be derived as a canonical distribution stemming 
either from the Tsallis ($S_T$) or from the R\'enyi-entropy ($S_R$) \cite{RENYI1,RENYI2,GE1,GE2,GE3,GE4}.
Both entropy formulas contain a $q=1+a$ parameter, and, from eqs.~(\ref{LOGCHAR}) and (\ref{CHAR_TSALLIS}), 
are in fact connected as
\be
S_R =K^{-1}(S_T)= \frac{1}{a} \ln \left(1+aS_T \right).
\ee{RENYI_TSALLIS}
In the $a \rightarrow 0$ limit $S_R$ coincides with Boltzmann's entropy formula.
The canonical energy distribution is derived from maximizing
\be
S_R - \beta \sum_i p_iE_i  - \alpha \sum_i p_i.
\ee{SRENYI_MAX}
Differentiation with respect to $p_i$ leads to
\be
p_i = A \left(\alpha+\beta E_i \right)^{-1/a},
\ee{RCANON_F}
with the proportionality constant
\be
A = e^{S_R} \, \left( \frac{a}{1-a} \right)^{-1/a}.
\ee{RENYI_PROP}
Here $\alpha$ and $\beta$ can be expressed by the other parameters according
to the normalization condition $\sum_i p_i = 1$ and the average energy.

In this paper we attempt to obtain an estimate for the $a=q-1$ parameter of the R\'enyi and
Tsallis entropy by investigating simple, static Schwarzschild black hole thermodynamics 
with maximizing the R\'enyi logarithm of the Bekenstein--Hawking horizon-entropy. As a result
we obtain an interesting thermodynamical behavior which is very similar to the one of a
black hole in anti-de Sitter (AdS) space \cite{HP}, with equation of state obtained by maximizing 
the original Bekenstein-Hawking entropy, proportional to the horizon area \cite{BH1,BH2,BH3}.
By an explicit comparison we demonstrate that the resulting thermodynamic temperature, $T=dE/dS$, 
follows an analogous dependence on the horizon radius in both systems. This offers a re-interpretation 
of the parameter $a$ in terms of the anti-de Sitter curvature parameter, $\lambda$.

For positive $a$ the resulting temperature has a minimum at a certain horizon radius, 
both smaller and larger black holes appear hotter and decay faster. By estimating this 
minimum-temperature radius based on the ground state energy of a semi-classical
string spanned over the diameter of the horizon, $2r=4E$ in Planck units, 
we consider the Bekenstein limit \cite{Bek} for quantum sized black holes.
Astronomical black holes on the other hand should posses huge heat capacity, so
correspondingly a much smaller magnitude for the parameter $a=1/C_0$. Accordingly, 
one can expect that the minimum-temperature radius is larger than the size of the 
largest observed black-hole horizon, so the Tsallis-R\'enyi parameter is
largely different for astronomical and for elementary objects.

All over this paper energy, mass and momentum is measured in multiples of the Planck mass, $M_P$,
while length and time in multiples of the Planck length, $L_P$. All equations relating
unlike quantities are to be supported by corresponding powers of $M_P$ and $L_P$.
The entropy is measured in units of the Boltzmann constant, $k_B$.
In a $c=1$, $k_B=1$ system one expresses the Planck constant as $\hbar = L_P M_P$ and 
Newton's gravity constant as $G=L_P / M_P$.

\section{Bekenstein-Hawking and R\'enyi entropy for black hole horizons}

The Bekenstein-Hawking entropy for simple (spherically symmetric, static) black hole horizons
can easily be obtained as follows. In general for a metric given by
\be
ds^2 = f(r) dt^2 - \frac{dr^2}{f(r)} - r^2 d\Omega^2 ,
\ee{METRIC}
with $t$ and $r$ being the time and radial coordinates for the far, static observer
and $d\Omega$ the two-dimensional surface angle, the corresponding Unruh temperature 
\cite{UNRUH} at the horizon becomes
\be
T = \frac{1}{4\pi} f'(r).
\ee{RAD_UNRUH}
Considering now that the internal energy is practically the mass of the black hole, 
one obtains the entropy according to Clausius' formula as an integral,
\be
S = 4\pi \int \frac{dM}{f'(r)}.
\ee{BH_ENTROPY}

This result can be written in a more elegant form by noting that the denominator,
$f'(r)$ -- to be evaluated at the condition $f(r)=0$ -- is a Jacobian for a
Dirac-delta constraint. Therefore the above BH-entropy equals to
\be
S = 4\pi \int\!\!\!\int \delta(f(r,M)) \, dr dM.
\ee{BH_SHELL}
This form reminds to a microcanonical shell in the phase space of the variables 
$r$ and $M=E$.

In the case of the Schwarzschild black hole solution one has
\be
f(r) = 1 - \frac{2M}{r},
\ee{SCHWARZ}
the horizon condition $f(r)=0$ is satisfied at $r=2M$
and the BH-entropy becomes
\be
S_{BH} = \pi r^2.
\ee{AREA_LAW}
Since the same result emerges for any $f(r,M)=1-2M/r-h(r)$, linear in $M$,
the entropy of such simple black hole horizons is always proportional to their area.
So the above ''area law'' is also valid for example for the Schwarzschild -- 
anti-de Sitter metric, too.

As it is known, the Bekenstein--Hawking result leads to an equation of state, 
$S(E)$ which describes an object with negative heat capacity:
\begin{eqnarray}\label{BH_EOS}
S_{BH} &=& 4\pi E^2, \nonumber\\
\frac{1}{T_{BH}} &=& S_{BH}'(E) = 8\pi E, \\
C_{BH} &=& \frac{-S_{BH}'^2(E)}{S_{BH}''(E)} = -8 \pi E^2\ .\nonumber
\end{eqnarray}

Recently, black hole thermodynamics has been actively investigated in the non-extensive 
framework \cite{Lan1,Lan2,Lan3,Lan4,Pav,Mad,Gour,Opp,Pes,Aran}, and the R\'enyi entropy 
formula has also been considered in this context \cite{BC,BMM}. In this paper, our novel 
approach is to regard the Hawking-Bekenstein entropy as a Tsallis entropy, 
and use its additive formal logarithm, the R\'enyi entropy (\ref{RENYI_TSALLIS}), 
for the canonical analysis instead. The essential difference is, that while the Tsallis
entropy is not additive for factorizing probabilities, the R\'enyi entropy is, and thus, 
the zeroth law of thermodynamics apply. Via equations (\ref{RENYI_TSALLIS}) and (\ref{BH_EOS}), 
one obtains the following result:
\begin{eqnarray}\label{RENYI_EOS}
S_R&=&\frac{1}{a}\ln\left(1+4\pi aE^2\right),\nonumber\\
\frac{1}{T_R}&=&S_R'(E)=\frac{8\pi E}{1+4\pi aE^2}, \\
C_R&=&\frac{-S_R'^2(E)}{S_R''(E)}=\frac{8\pi E^2}{4\pi aE^2-1}\nonumber .
\end{eqnarray}

\section{Temperature and horizon radius}

By analyzing the thermodynamics given by the equations in (\ref{RENYI_EOS}), one realizes 
that it is surprisingly similar to the one of a black hole in AdS space \cite{HP}.
Consequently we now obtain simultaneously the R\'enyi-entropy based thermodynamics 
of the simple Schwarzschild black hole (discussed above) and the original Bekenstein-Hawking 
entropy based thermodynamics of a black hole in anti-de Sitter space. From now on we use the
notation $E=M(r)$ with $r$ being the horizon radius.

In the first case we have the metric factor
$f(r)=1-2M/r$ from which $M(r)=r/2$ on the horizon. The considered entropy is
the R\'enyi one, $S=\frac{1}{a}\ln(1+a\pi r^2)$. The temperature becomes
\be
T(r) = \frac{M'(r)}{S'(r)} = \frac{1}{4\pi r} + \frac{a}{4} r.
\ee{BH_HORIZON_TEMP}
In the second case the metric factor is modified to $f(r)=1-2M/r+\lambda r^2/2$
and 
\be
M(r)= \frac{r}{2} \left(1 + \lambda \frac{r^2}{2} \right).
\ee{ADS_MASS_HORIZON}
Its radial derivative becomes
\be
M'(r) = \frac{1}{2} + \lambda \frac{3r^2}{4}
\ee{ADS_MASS_DERIV}
giving rise to the temperature
\be
T(r) =  \frac{M'(r)}{S'(r)} = \frac{1}{4\pi r} + \frac{3\lambda}{8\pi} r.
\ee{ADS_TEMP_HORIZ}
This intriguing coincidence motivates to consider a R\'enyi entropy with
$a=3\lambda/2\pi$ simulating the anti-de Sitter Boltzmannian thermodynamics 
for positive parameter values. In both cases there is a minimal temperature. 
The radius where this occurs satisfies $T'(r_0)=0$, in the R\'enyi case 
$q=1+a=1+1/\pi r_0^2$. At this radius the heat capacity has a pole: 
\be
C(r)=\frac{dM}{dT}=\frac{M'(r)}{T'(r)}=\frac{2\pi r_0^2 r^2}{r^2-r_0^2}\ .
\ee{HEAT_CAP_RENYI}
The $q$-parameter for an ideal gas on the other hand
is given as $q=1+1/C_0$. This means that the characteristic ideal-gas-equivalent 
heat capacity, coded in this parameter, belongs to a positive $C_0=\pi r_0^2 = 
S_{{\rm BH}}(r_0)$, actually not occurring in $C(r)$, since it stays over the 
asymptotic value $2\pi r_0^2$ when positive.

\section{Black hole evaporation}

Classical Schwarzschild black hole horizons are bad heat containers
due to their negative heat capacity. In the R\'enyi thermodynamics,
or equivalently with the AdS metric, the heat capacity, $C=dM/dT=M'(r)/T'(r)$ 
is negative only for $r < r_0$, since in both cases $M(r)$ is monotonic rising,
while $T(r)$ is falling for $r<r_0$ black hole horizons. At $r=r_0$ the heat 
capacity diverges, a pole occurs. Above $r_0$ the $T(r)$ curve is rising again 
featuring a positive heat-capacity system.

\begin{figure}
\begin{center}
\includegraphics[width=0.4\textwidth, angle=-90]{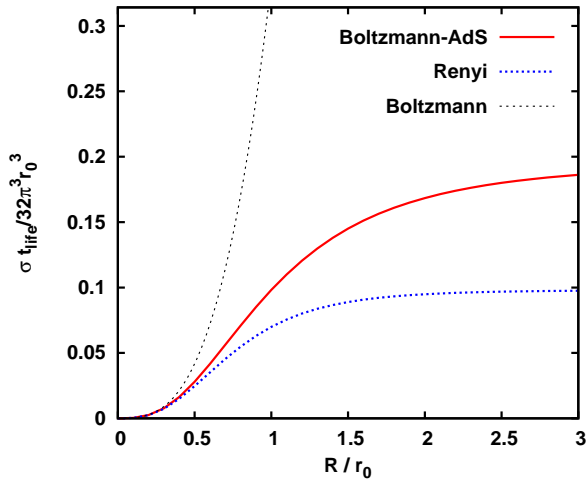}
\end{center}
\caption{
\label{Fig1}
The scaled lifetime of spherical static black holes against black-body radiation
in the R\'enyi-Schwarzschild and in the Boltzmann-AdS scenarios as functions of
the initial radius scaled by the radius where the temperature is minimal, $R/r_0$.
}
\end{figure}

In the simplest evaporation scenario, the black-body radiation, belonging to 
the temperature $T(r)$, consumes the energy, $M(r)$, of the black hole. Using 
the Stefan-Boltzmann formula for the energy flux, one can model the energy loss 
as
\be
\dot{M} = M'(r) \dot{r} = - 4\pi r^2 \, \sigma T^4(r).
\ee{STEFANBOL}
Since we have obtained the same functional form $T(r)$ for the R\'enyi-Schwarzschild and for
the Boltzmann-AdS case, only the $M'(r)$ factor distinguishes these cases.
Due to the positivity of all factors, it is obvious that the lowest decay rate occurs
roughly at $r=r_0$, so adiabatically radiating simple black holes spend the longest part of 
their life time near to this radius. The finite decay time, during which an initial radius, 
$R$, shrinks to zero, can be analytically obtained in both cases. It is given by
\be
t_{{\rm life}}(R) = \frac{(4\pi r_0)^3}{2\sigma} \, \int_0^{R/r_0}\!\frac{x^2}{(1+x^2)^n} \, dx
\ee{LIFETIME}
with $n=3$ for the AdS and $n=4$ for the R\'enyi scenario.
The analytic results of the integrations are
\be
t_{{\rm life}}^{{\rm AdS}}(R) = \frac{4\pi^3r_0^3}{\sigma} \, 
\left[ {\mathrm{\arctan}}(x) + \frac{x(x^2-1)}{(x^2+1)^2} \right]
\ee{INTEGRAL_LIFE_ADS}
for $n=3$ (AdS), and
\be
\!\!\!\!\!\!\!\!\!\!\!\!
t_{{\rm life}}^{{\rm Renyi}}(R) = \frac{2\pi^3r_0^3}{\sigma} \, 
\left[ {\mathrm {\arctan}}(x) + \frac{x(x^4+\frac{8}{3}x^2-1)}{(x^2+1)^3} \right]
\ee{INTEGRAL_LIFE_RENYI}
for $n=4$ (R\'enyi). Here we used the scaling notation $x=R/r_0$.
For small initial radii $R \ll r_0$, both evolutions are well approximated by
the classical Schwarzschild result
\be
t_{{\rm life}}^{{\rm class}}(R) = \frac{32\pi^3R^3}{3\sigma}. 
\ee{INTEGRAL_LIFE_SMALL}
However, very large black holes do not live forever in this approach:
they also have a finite decay time. The maximal lifetime differs by a factor of two
in these cases: 
\be
t_{{\rm life}}^{{\rm Renyi}}(\infty) = \frac{\pi^4r_0^3}{\sigma} 
\ee{INTEGRAL_LIFE_INFINITE_RENYI}
for the R\'enyi thermodynamics, while
\be
t_{{\rm life}}^{{\rm AdS}}(\infty) = \frac{2\pi^4r_0^3}{\sigma} 
\ee{INTEGRAL_LIFE_INFINITE_ADS}
in the conventional thermodynamics of an AdS black hole.
Figure \ref{Fig1} plots the entire evolution of the horizon radii.

\section{A semi-classical estimate for the $q$-parameter}

In particle physics, one conjectures small black holes, so $q$ may deviate from
unity appreciably. For astronomical black holes it is probably practically unobservable
(the maximal lifetime depending on the effective heat-capacity, $C(r_1)$, to which
$q$ has been fitted to). In the followings we deal with an estimate for the horizon 
radius at the minimal temperature, $r_0$, for near-quantum black holes.

If the energy of such a state, $E_0=M(r_0)$, lied below or around the quantum mechanical 
ground state energy, then we have reached the edge of the classical theory discussed so far.
Of course, in the absence of a functioning quantum theory of gravity, one can only
have an estimate for this limit from semi-classical considerations. The simplest of which
is regarding the energy of a string of the length of the diameter, $2r_0$, having a wave
number and frequency $\omega = k = \pi/2r_0$ (the sinus wave with no intermediate nodes
and massless excitations). The corresponding ground state quantum oscillator energy is 
required to be greater than the classical black hole energy at the turning minimum point 
of the temperature:
\be
\frac{\omega}{2} = \frac{\pi}{4r_0} \ge E_0 = \frac{r_0}{2}.
\ee{GROUND_STATE}
This translates to the condition $r_0^2 \le \pi/2$.
Utilizing now the relation $a=1/(\pi r_0^2)$  one concludes that 
\be
q = 1 + a \ge 1 +  \frac{2}{\pi^2}.
\ee{qESTIMATE}
It is amazing to realize that this estimate, $q\approx 1.2026$, is how well approximated
by cosmic ray observations ($q=11/9$) \cite{BECK1,BECK2} and by the quark coalescence fit to
RHIC hadron transverse momentum spectra ($q \approx 1.2$) \cite{SQM2008}.

\section{Summary}

By interpreting the Bekenstein-Hawking entropy as a non-extensive Tsallis
entropy of Schwarzschild black hole horizons, and by considering 
their equation of state based on the R\'enyi entropy, one obtains a temperature 
minimum at a given horizon radius $r_0$. This feature (and the whole $T(r)$ curve) 
is of the same form as the result from a black hole in anti-de Sitter space by using 
the Boltzmann-Gibbs entropy. Requiring that the energy at this 
point, $E_0=M(r_0)$, is above a semi-classical ground state of a string stretched
over the diameter $2r_0$ of the horizon, amazingly a Bekenstein bound for the $q$-parameter 
of R\'enyi's and Tsallis' entropy formulas arises. Findings in relativistic heavy ion 
collisions and in cosmic ray observations are characterized by a $q$-value surprisingly 
close to this bound.

Recently, black holes in AdS spacetimes seem to be relevant through the Maldacena conjecture 
\cite{MALCADENA}, and it might be an interesting question whether our findings may be 
better understood or reinterpreted from the holographic gauge/gravity correspondence point 
of view \cite{WITTEN,KLEBA}.

Although it is well known that micro black holes can not form in high energy particle and heavy ion 
collisions in 3-dimensions, nevertheless, due to the extreme deceleration by the stopping, a Rindler horizon 
may occur for the newly produced hadrons. This can be, in general, the origin of thermal looking spectra 
\cite{BiroSchramGyulassy} as also suggested using different argumentation by Kharzeev and Satz earlier 
\cite{KHARZEEV,SatzKharzeev}. This mechanism can well be the reason behind the leading non-extensive 
effect in the power-law spectra, consistent with a statistical, constituent quark matter hadronization 
picture.

In addition, it is also important to emphasize that for producing features analogue to the 
Hawking radiation in the spectrum, no black hole formation is needed in the classical sense. The 
formation of an event horizon suffices. As an example, it has been recently reported about ultra-short 
laser pulse experiments, that among other known radiations, a Hawking radiation of photons has also been 
produced \cite{LASER}.

Finally we would like to emphasize that if experimental findings would vastly violate
the bound derived in eq.~(\ref{qESTIMATE}) on the quark level, -- mesonic spectra
showing half, while baryonic spectra one third of the $a=q-1$ quark level value --
then it may be a strong argument against speculations on the presence of event
horizon related phenomena in high energy particle collisions.

\section*{Acknowledgement}
T.S.B.~gratefully acknowledges discussions with Profs. C. Tsallis, 
B.~M\"uller and A.~Jakov\'ac. V.G.Cz.~has benefited from a discussion with Sameer Murthy.
This work has been supported by Hungarian National Research Fund OTKA grant K104260, by 
a bilateral Hun\-ga\-ri\-an-South-African project NIH TET\_10-1\_2011-0061 and ZA-15/2009, 
by the T\'AMOP 4.2.1./B-09/1/KONV-2010-0007 project, co-financed by the European Union 
and European Social Fund, and was also partly supported by the Helmholtz International 
Center for FAIR within the framework of the LOEWE program (LandesOffensive zur Entwicklung 
Wissenschaftlich-\"Okonomischer Exzellenz) launched by the State of Hesse. The research 
leading to these results has also received funding from the European Union Seventh Framework 
Programme (FP7/2007-2013) under grant agreement n$^{\circ}$ PCOFUND-GA-2009-246542 and from 
the Foundation for Science and Technology of Portugal.

\end{document}